# Quantitative Damping Calculation and Compensation Method for Global Stability Improvement of Inverter-Based Systems

Yang Li, Zenghui Zheng, Xiangyang Wu, Jiayong Li, *Member, IEEE*, Wei Wang, Qiang Zeng, Zhikang Shuai, *Senior Member, IEEE*

*Abstract*- Small-signal stability issues-induced broadband oscillations pose significant threats to the secure operation of multi-inverter systems, attracting extensive research attention. Researches revealed that system instability is led by the lacking of positive damping, yet it has not been clearly specified how much the exact amount of damping compensation required to sufficiently ensure system global stability. This paper presents a feasible solution for quantitative damping calculation and compensation to enhance the global stability of inverter-based systems. First, based on the system nodal admittance model, a quantitative damping calculation algorithm is presented, which can suggest the required damping compensation as well as compensation location for sufficient stability improvement. Then, we propose a specific AD with output current feedforward control strategy, which make the AD be quasi-pure resistive and can effectively enhance system damping efficiency. Finally, a testing system with three inverters is used as case study, showing that the proposed method provides a promising solution to efficiently enhance the global stability improvement of inverter-based systems. Simulations and experiments validate the proposed method.

## I. INTRODUCTION

WITH the large-scale integration new energy such as wind or solar via power electronic inverters, instability issues caused by the impedance interactions among inverters and grid are becoming increasingly serious, posing threats to the normal operation of power systems [1]-[2]. Researches have shown the negative damping effect resulted by phase locked loop (PLL) or time delay as well as the interaction resonance between inductive and capacitive components is the fundamental cause of system instability [3]-[5]. Thus effective positive damping compensation is of great significance for stability improvement of inverter-based systems.

Impedance-based techniques are powerful and commonly used for stability assessment of inverter-based systems [5]-[12]. Such kind of methodologies for inverter-based systems employs port impedance or admittance of apparatus and grid interfaces, circumventing state-space modeling that demands full-system parameter identification [7]. This merit enables stability investigation of grey/black-box systems, where the port impedance/admittance of inverters can be acquired by impedance measurement [11], [12]. Furthermore, the artificial intelligence techniques can be employed to obtain accurate impedance/admittance at multi-operating points with limited data inputs [13]. Ref. [8] presented a positive-net-damping stability criterion to analyze electrical instability, which can characterize the frequency of closed-loop oscillatory modes and reveal the damping of these modes. Ref. [10] proposed stability analysis and location optimization method for multi-inverter power systems, and used the participation factors of the return-ratio matrix's critical eigenvalues to reveal global stability feature of the system. Ref. [11] also used participation factors for instability root-cause tracing of power system stability, which can identify "trouble makers" of the system. These research studies demonstrate the superiority of impedance-based analysis methods in investigating the small-signal stability of multi-inverter systems.

The primary objective of stability assessment is to ensure system stabilization. Refs. [6] and [7] calculated eigenvalue sensitivity of system impedance model with respect to control and component parameters, to assess then that have dominant influences on the stability can be identified and re-tuned for stability improvement. Also, one can reshape impedances of inverters by means of modifying inverter control strategies to provide positive damping and improve system stability margin [14]-[16]. However, these approaches exhibit fundamental limitations when addressing systems equipped with controllability-deficient black-box inverters. To solve this problem, active dampers (ADs) are used and recent of great interest. Such kind of ADs can either be integrated with battery energy storage system (BESS) [17], or function as a standalone damping device [18], [19]. Ref. [17] proposed an enhancing grid stiffness control strategy of STATCOM/BESS for improving damping in wind farm connected to weak grid. In [18], an adaptive active damper was presented to cope with the instability issues of the *LCL*-type gird-connected inverter. In [19], a low voltage active damper (LVAD) was proposed for the stability enhancement of microgrids with unknown-parameter inverters, and parameter design of the damper are discussed. However, it has not been clearly specified how much the exact amount of damping compensation required to sufficiently ensure system stability.

Facing the integration of a large number of inverters into complicated grid conditions such as weak grids and resonance caused by interactions between inductive and capacitive components [3], [5], the hazards of instability-induced oscillations can rapidly propagate throughout the entire system. To address such issues effectively, two key points deserve further attention:

1) For multi-inverter integrated systems, it is essential to quantitatively calculate the required damping compensation for global stability improvement, which is supposed to provide the guidance of design and installing location optimization for damping devices.
2) Damping devices shall possess preferable impedance/admittance characteristics that will be applicable to complicated grid conditions and can effectively achieve broadband damping compensation in wide frequency range.

This paper focuses on the above key points and presents a feasible solution for quantitative damping calculation and compensation to enhance the global stability of inverter-based systems. The rest of this paper is organized as follows: Section II presents a quantitative damping calculation algorithm for stability improvement of multi-inverter systems based on the whole system nodal admittance model. Section III proposes a specific AD that is quasi-pure resistive and can effectively enhance system damping efficiency. In Section IV, a testing system is used as case study to validate the proposed method, and Section V presents the experimental validations. Section VI summarizes conclusions.

## II. MODELING AND QUANTITATIVE DAMPING CALCULATION OF INVERTER-BASED SYSTEMS

### A. System Modeling and Stability Criterion

Without loss of generality, as depicted in Fig. 1, assume that $m$ inverters are connected to the three-phase system with $n$ nodes. $V_i$ and $I_i$ represents the $i$-th node voltage and injection current, respectively. $Y_{eqi}$ represents the $i$-th inverter output admittance. The injection current $I_i$ composes of the Norton equivalent current source of the $i$-th inverter (i.e., $I_{Si}$) and equivalent current source of the grid (if exists, e.g., $I_g$). The system can be modeled by either dq frame impedances or sequence impedances, and this paper uses the former. All inverter output admittances are established in a global dq frame.

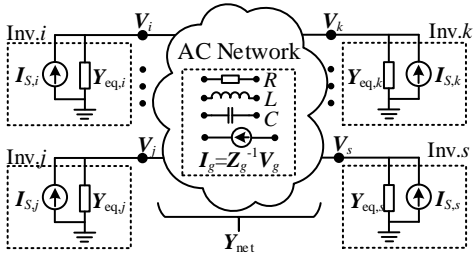

Fig. 1. Schematic diagram of inverter-based system.

The node voltage can be obtained by
$$V_{\text{nod}} = Y_{\text{nod}}^{-1} I_{\text{nod}} \quad (1)$$
where $V_{\text{nod}}=[V_1, \ldots, V_i, \ldots, V_n]^T$ represents the node voltage vector, $V_i=[V_{id}, V_{iq}]^T$; $I_{\text{nod}}=[I_1, \ldots, I_i, \ldots, I_m]^T$ represents the node injection current vector, $I_i=[I_{id}, I_{iq}]^T$; $Y_{\text{nod}}$ represents the node admittance matrix of the whole system. $Y_{\text{nod}}$ composes of $Y_{\text{net}}$ (node admittance matrix of AC network) and $Y_{\text{inv}}$ (node admittance matrix of inverters). $Y_{\text{inv}}=\text{diag}(\ldots, Y_{eqi}, \ldots, Y_{eqj}, \ldots)$.

Note that each inverter is self-stable (which is the basic requirement for the inverter manufacturers) thus $Y_{\text{inv}}$ has no right half plane pole (RHP), and $Y_{\text{net}}$ that contains only passive components also has no RHP. Then based on the impedance-sum-type criterion (derived from Cauchy's theorem) [9], the system's stability is determined by whether the Nyquist trajectory of eigenvalues of $Y_{\text{nod}}$ encircle point (0, 0) or not.

The eigenvalue Nyquist trajectory-based stability criterion can also be represented through the frequency-dependent parametric trajectories formed by the real and imaginary axes projections of eigenvalues of $Y_{\text{nod}}$. As illustrated in Fig. 2, assuming the Nyquist contour of eigenvalue $\lambda$ undergoes encirclement around the origin (0, 0) and intersects the real axis at crossover frequency $f_{cr}$ (designated as intersection point A), the subsequent parametric analysis reveals: when projecting $\lambda$'s real and imaginary components as frequency-dependent functions, the resulting loci demonstrate Im[$\lambda$]=0 with Re[$\lambda$]<0 at $f_{cr}$. In other words, the closed-loop system in (1) is asymptotically stable if at the frequency of Im[$\lambda$] equal to 0, Re[$\lambda$] is positive, i.e., Im[$\lambda$]($f_{cr}$)=0 and Re[$\lambda$]($f_{cr}$)>0. In fact, this is an extension of Positive-Net-Damping Stability Criterion [8] applying to a multiple-input multiple-output system, and it can be called Generalized Positive-Net-Damping Stability Criterion (GPNDSC).

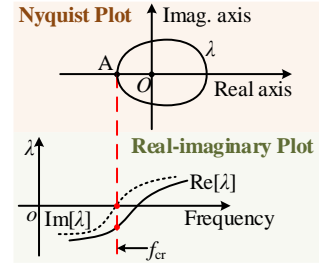

Fig. 2. Nyquist plot of eigenvalue and its real and imaginary parts as functions of frequency.

It is worth pointing out that, at crossover frequency $f_{cr}$, the zero imaginary part of critical eigenvalues indicates the system's oscillation potential, while the negative real part implies that negative damping sustains oscillatory modes by continuously amplifying their amplitudes.

The merit of using node admittance matrix $Y_{\text{nod}}$ based on GPNDSC to assess the stability of inverter-based systems is to avoid formulating complicated return ratio matrix, which is very engineering-friendly for complex structured systems. The eigenvalues those not satisfy GPNDSC will be considered as the critical ones, and tend to be the focuses in stability improvement. Certainly, the real parts of critical eigenvalues at crossover frequences are the main concerns.

### B. Quantitative Requires Damping Compensation Calculation

The derivative of the critical eigenvalue $\lambda_k$ of node admittance matrix $Y_{\text{nod}}$ with respect to a certain parameter $\alpha$ can be expressed as

$$\frac{\partial \lambda_k(\alpha)}{\partial \alpha}=u_k^T \frac{\partial Y_{\text{nod}}(\alpha)}{\partial \alpha} w_k = \begin{bmatrix} \cdots & u_{ki} & \cdots \end{bmatrix} \begin{bmatrix} 0 & \cdots & 0 \\ \vdots & \frac{\partial(Y_{ij}+\alpha)}{\partial \alpha} & \vdots \\ 0 & \cdots & 0 \end{bmatrix} \begin{bmatrix} \vdots \\ w_{jk} \\ \vdots \end{bmatrix} = u_{ki} w_{jk} \quad (2)$$

where $u_k=[u_{k1}, u_{k2}, \ldots, u_{k(2m)}]$ and $w_k=[w_{k1}, w_{k2}, \ldots, w_{k(2m)}]^T$ are left and right eigenvectors corresponding to $\lambda_k$, respectively.

Let $\alpha = G + jB$, then one can get that

$$\frac{\partial \lambda_k}{\partial G} = \frac{\partial \lambda_k}{\partial \alpha} \cdot \frac{\partial \alpha}{\partial G} = u_{ki} w_{jk} = S_{re} + jS_{im}$$
$$\frac{\partial \lambda_k}{\partial B} = \frac{\partial \lambda_k}{\partial \alpha} \cdot \frac{\partial \alpha}{\partial B} = j \cdot u_{ki} w_{jk} = -S_{im} + jS_{re}$$
(3)

Ep. (3) indicates that the real part of $\partial \lambda_k / \partial \alpha$, i.e., $S_{re}$, represents the derivatives of the real and imaginary parts of $\lambda_k$ with respect to the real and imaginary parts of $\alpha$, respectively; while the imaginary part of $\partial \lambda_k / \partial \alpha$, i.e., $S_{im}$, represents cross-derivatives of the real and imaginary parts of $\lambda_k$ with respect to the imaginary and real parts of $\alpha$, respectively.

According to (2), a minor variable disturbance $\Delta \alpha = \Delta G + j \Delta B$ will cause the increment of the critical eigenvalue $\lambda_k$ to be

$$\Delta \lambda_k = \Delta \alpha \cdot \frac{\partial \lambda_k(\alpha)}{\partial \alpha} = \Delta \alpha \cdot u_{ki} w_{jk} \quad (4)$$

To improve the system's stability, the real part of the critical eigenvalue $\lambda_k$ at crossover frequence must be elevated to a certain value greater than $\varepsilon$ (where $\varepsilon > 0$) through stability augmentation measures, establishing a positive damping margin in the system's frequency response, namely

$$\text{Re}[\lambda_k] + \text{Re}[\Delta \lambda_k]\big|_{f=f_{cr}} > \varepsilon \quad (5)$$

Then the desired damping compensation for stability improvement can be calculated as

$$\text{Re}[\Delta \lambda_k]\big|_{f=f_{cr}} = \text{Re}\left[\Delta \alpha \cdot u_{ki} w_{jk}\right]\big|_{f=f_{cr}} > \varepsilon - \text{Re}[\lambda_k]\big|_{f=f_{cr}} \quad (6)$$

According to (6), the required $\Delta \alpha$ to meet the demand of damping compensation for stability improvement can be determined. However, considering that $\partial \lambda_k(\alpha)/\partial \alpha$ could be vary with change of $\alpha$, a step-by-step procedure should be implemented to obtain a sufficient value of $\alpha$. The detailed procedure for calculating $\alpha$ to meet the demand of damping compensation is given as follows.

Step 1 Collect system data, which includes information of network structure, line and grid impedances, inverter admittances, to obtain node admittance matrix $\boldsymbol{Y}_{nod}$.

Step 2 Identify critical eigenvalues, and then calculate the sensitivity coefficients by (2).

Step 3 Calculate $\Delta \lambda_k$ by (4) to obtain the desired damping compensation $\text{Re}[\Delta \lambda_k]$ at crossover frequency with a small increment $\Delta \alpha$.

Step 4 Implement comparison in (6), if it not satisfies, add increment $\Delta \alpha$ to $\alpha$ and repeat Step 2 – Step 4 until (6) is satisfied.

Pseudocode for the above procedure is presented as Algorithm 1.

| **Algorithm 1**: calculating $\alpha$ to meet the demand of damping compensation |
|---|
| 1:     collect system data to obtain $\boldsymbol{Y}_{nod}$ |
| 2:     initialize $\Delta \lambda_k = 0$, $\alpha = 0$ |
| 3:     **while** $\text{Re}[\Delta \lambda_k] > \varepsilon - \text{Re}[\lambda_k]$ **do** |
| 4:         identify critical eigenvalues with the consideration of $\alpha$ |
| 5:         calculate $\partial \lambda_k(\alpha)/\partial \alpha$ by (2) at $f_{cr}$ |
| 6:         calculate $\Delta \lambda_k = \Delta \lambda_k + \Delta \alpha \cdot u_{ki} w_{jk}$ |
| 7:         update $\alpha = \alpha + \Delta \alpha$ |
| 8:     **end** |
| 9:     **end** |

Note that an unstable system may have more than one critical eigenvalue, so all critical eigenvalues should be sufficiently compensated at their crossover frequencies. Besides, the optimal placement of damping compensation devices requires systematic evaluation of compensation efficacy, with priority given to locations exhibiting higher sensitivity coefficient $\partial \lambda_k(\alpha)/\partial \alpha$. Also note that the proposed algorithm of damping calculation based on sensitivity coefficient can be applicable to the so-called return-ratio matrix $\boldsymbol{Y}_{eq}\boldsymbol{Z}_m$ [10] according to the chain rule of differentiation.

## III. QUASI-PURE RESISTIVE ACTIVE DAMPER FOR STABILITY IMPROVEMENT OF INVERTER-BASED SYSTEMS

### A. Damping Compensation by Active Damper

The intrinsic mechanism underlying damper-enhanced system stability arises from admittance matrix modification induced by output admittance of AD integrated into the power grid, which subsequently alters system damping characteristics through reshaping critical eigenvalues.

When an AD connected to a node of the network, e.g., the $i$-th node, neglecting the cross couplings between d- and q-axis in AD admittance matrix (a specialized AD whose dq and qd admittance can be neglected will be discussed in next part), the elements $Y_{(2i-1)(2i-1)}$ and $Y_{(2i)(2i)}$ in $\boldsymbol{Y}_{nod}$ will be added the AD's admittance $Y_{ad}$, then based on (2)-(4), one can get

$$\Delta \lambda_k = Y_{ad} \cdot \left( u_{k(2i-1)} w_{(2i-1)k} + u_{k(2i)} w_{(2i)k} \right) = Y_{ad} \cdot K_C \quad (7)$$

Then the required damping compensation from $Y_{ad}$ for stability improvement can be obtained by Algorithm 1. $K_C$ is defined as compensation coefficient of AD.

Two key points need to be discussed sequentially. Firstly, $Y_{ad}$ and $K_C$ both have real and imaginary parts, thus $\Delta \lambda_k$ will also have real and imaginary parts. Consequently, the imaginary part of critical eigenvalue will be affected, causing its crossover frequency $f_{cr}$ to change ($f_{cr}$ may increase or decrease). When crossover frequency exhibits significant variation, it renders challenging the precise augmentation of $\text{Re}[\Delta \lambda_k]$ at crossover frequency through the calculation of $Y_{ad}$ in (7) for stability improvement. Secondly, the real and imaginary parts of $Y_{ad}$ both involve in augmentation of $\text{Re}[\Delta \lambda_k]$, which amplifies the complexity inherent in Algorithm 1 to obtain appropriate $Y_{ad}$. The above two key points will inevitably introduce computational complexities in damping compensation calculation and pose challenges for damper design.

We will investigate how the real and imaginary parts of AD output admittance influence critical eigenvalue characteristics from a perspective of mathematical derivation and physical mechanism. Let us consider a specific scenario: the inverter-based system has one node, where inverters and grid impedance are parallel connected. For this scenario, its critical eigenvalue $\lambda_k$ satisfies

$$\left| \lambda_k \boldsymbol{E} - \boldsymbol{Y}_{nod} \right| = 0 \quad (8)$$

where $\boldsymbol{E}$ represents a $2 \times 2$ unit matrix.

When AD is shunted into this node, the node admittance matrix of PCC becomes $\boldsymbol{Y}_{nod} + [Y_{ad}, 0; 0, Y_{ad}]$, then one can get

$$\left| (\lambda_k - Y_{ad}) \boldsymbol{E} - \boldsymbol{Y}_{nod} \right| = 0 \quad (9)$$

The mathematical derivation in (8) reveals that critical eigenvalue of $Y_{nod}+[Y_{ad}, 0; 0, Y_{ad}]$ is $\lambda_k'=\lambda_k+Y_{ad}$. It means that for such specific scenario, the real and imaginary parts of $Y_{ad}$ superimpose on those of $\lambda_k$, respectively, where no cross-coupling effect exists between real and imaginary parts. This can be interpreted from the perspective of physical mechanism i.e., the system's oscillatory modes (imaginary parts of critical eigenvalues and crossover frequencies) will be mainly affected by the susceptance of $Y_{ad}$ (its imaginary part), while the damping will be mainly affected by the conductance of $Y_{ad}$ (its real part). This physically interpretable conclusion can be systematically generalized to structurally complex power systems, i.e., the real parts of critical eigenvalues are mainly affected by real part of $Y_{ad}$, while imaginary parts of critical eigenvalues are mainly affected by imaginary part of $Y_{ad}$, and the cross-coupling effect between real and imaginary parts will be relatively small. This can greatly reduce algorithm complexity of Algorithm 1 to obtain an appropriate $Y_{ad}$ for stability improvement. Also, it indicates that a purely resistive AD is desired for effectively enhancing system damping efficiency in critical eigenvalue trajectory optimization.

*B. Proposed Active Damper and Its Control Strategy*

Leveraging the control flexibility of power electronic converters, this paper proposes a quasi-pure resistive damper to efficiently compensate negative damping in inverter-based systems. The topology and control block diagram proposed AD is shown in Fig. 3. The proposed AD's control strategy includes fundamental and damping control loops. To achieve wide-frequency damping compensation across the entire grid spectrum, the damper shall adopt an ultra-high switching frequency (e.g., ≥40 kHz) to enable effective modulation of high-frequency components in the carrier wave.

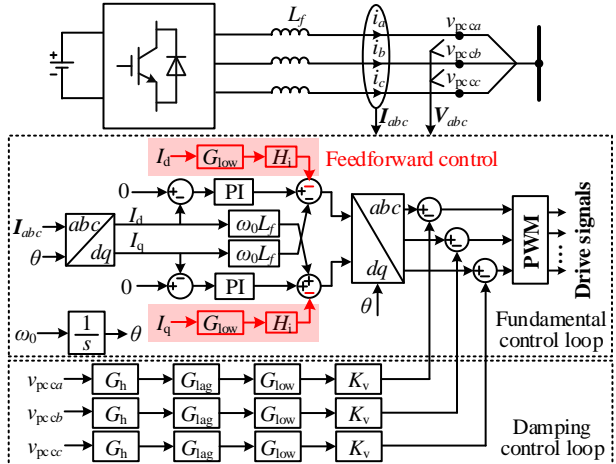

Fig. 3. Topology and control block diagram of the proposed AD.

In fundamental control loop, d- and q-axis currents are controlled to zero in dq frame. Normally since AD does not engage in fundamental power transmission, PCC voltage synchronization phase angle is not required for its operational implementation. Thus, synchronous rotation phase angle $\theta$ is generated by integral of fundamental frequency $\omega_0$, which can avoid the negative effects of PLL in weak grid [3]. In addition, to preclude the outer voltage control loop from affecting damping performance in relatively low-frequency bands [20], the proposed AD has no an outer voltage control loop. Its DC-side voltage is supplied by BESS, ensuring independent damping operation from DC voltage regulation dynamics. Under grid contingency scenarios, the damper can dynamically switch to traditional control modes, enabling coordinated participation of BESS in grid peak shaving and valley filling operations [21].

In AD's damping control loop, non-fundamental component of PCC voltage is extracted by a notch filter $G_h$, and a lag filter $G_{lag}$ is used to compensate the led phase introduced by $G_h$ in the vicinity of the upper sideband of the fundamental frequency. A low pass filter $G_{low}$ with relatively high cut-off frequency is used to prevent high-frequency interference. $K_v$ is the compensation coefficient. The specific expressions of $G_h$, $G_{lag}$, and $G_{low}$ are given in (10)-(12).

$$G_h(s) = \frac{s^2 + \omega_0^2}{s^2 + 2\xi\omega_0 s + \omega_0^2} \quad (10)$$

$$G_{lag}(s) = \frac{\tau s + 1}{\beta\tau s + 1} \quad (11)$$

$$G_{low}(s) = \frac{\omega_{low}}{s + \omega_{low}} \quad (12)$$

where $\xi$ is damping ratio of notch filter; $\tau$ is time constant of lag filter, and $\beta$ is ratio of zero locations to pole locations in the s-plane; $\omega_{low}$ is cut-off frequency of low pass filter.

Without outer voltage control loop and PLL, d- and q- axis of the proposed AD can be decoupled. Block diagram of the current control loop of the proposed AD is demonstrated in Fig. 4, where $Z_f(s)=sL_f$ denotes filter impedance of AD, and $G_d(s)=\exp(sT_d)$ is the time delay introduced by the digital control ($T_d=1.5/f_s$, and $f_s$ is the sampling frequency).

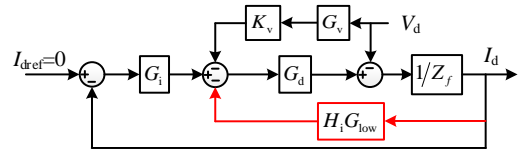

Fig. 4. Block diagram of the current control loop of proposed AD in d-axis.

In the proposed AD, as depicted in Fig. 3, the feedforward control of output current strategy is employed to reshape the impedance characteristics of the filter inductor $L_f$, which modifies the system's frequency response to exhibit low-pass filter behavior in relative high frequency bands, thereby filter impedance of AD displays quasi-pure resistive characteristics. To make

$$\frac{1}{sL_f + H_i(s)} = G\frac{\omega_c}{s + \omega_c} \quad (13)$$

where $H_i(s)$ is the feedforward control transfer function, $\omega_c$ is cut-off frequency of the intended low pass filter, and $G$ is the gain of this low pass filter.

Then it can get feedforward control transfer function as

$$H_i(s) = \left(\frac{1}{G\omega_c} - L_f\right)s + \frac{1}{G} \quad (14)$$

The expression of $H_i(s)$ includes a differential term $s$, thus the low pass filter $G_{low}$ is also used in current feedforward control strategy to prevent high-frequency interference. Based on the block diagram of the current control loop of proposed AD shown in Fig. 4, it can get the AD's admittance $Y_{ad}$ as

$$Y_{ad}(s) = -\frac{I_d}{V_d} = \frac{1 + K_v G_v G_d}{sL_f + H_i G_{low} G_d + G_i G_d} \quad (15)$$

where $G_v(s) = G_h(s+j\omega_0)G_{lag}(s+j\omega_0)G_{low}(s+j\omega_0)$.

Fig. 5 demonstrates the real and imaginary parts of $Y_{ad}$ of the proposed AD based on the theoretical model in (5). It can be seen that the imaginary part of $Y_{ad}$ is near zero in the frequency range within 2 kHz, while its real part has a certain positive value, which means the proposed AD is quasi-pure resistive in the frequency range within 2 kHz. Fig. 5 also shows the measurement data of the real and imaginary parts of dd and qq admittance of the proposed AD in dq frame, which illustrates the consistency of the theoretical model and measurement data.

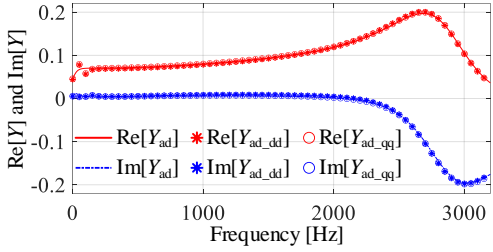

Fig. 5. Real and imaginary parts of $Y_{ad}$ of the proposed AD.

Fig. 6 shows the absolute value of the ratio between the imaginary and real parts of admittance of the proposed AD and the traditional one (which, for example, has no the proposed feedforward control strategy, i.e., $H_i$=0). It can be found that the real part of admittance of the proposed AD is substantially ten times greater than its imaginary part in the frequency range within 2 kHz, while the imaginary part of admittance of the traditional AD is greater than its real part with the increase of frequency (its admittance characteristic curves can be found in Fig. 12).

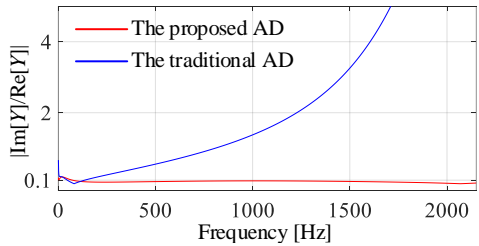

Fig. 6. |Im[Y]/Re[Y]| of the proposed AD and the traditional one.

Fig. 7 shows the clusters of the real part (the solid lines) and imaginary part (the dash lines) of admittance of the proposed AD with parameters varying. Herein, three parameters, i.e., $L_f$, $G$, and $K_v$, are taken as the examples. It can be found that, smaller $L_f$ can provide quasi-pure resistive $Y_{ad}$ in wider frequency range; smaller $G$ can also provide quasi-pure resistive $Y_{ad}$ in wider frequency range, but it will decrease Re[$Y_{ad}$], whereas larger $K_v$ can provide larger Re[$Y_{ad}$]. Based on the proposed AD's admittance curve cluster of, appropriate parameters can be configured to meet the system's damping compensation requirements.

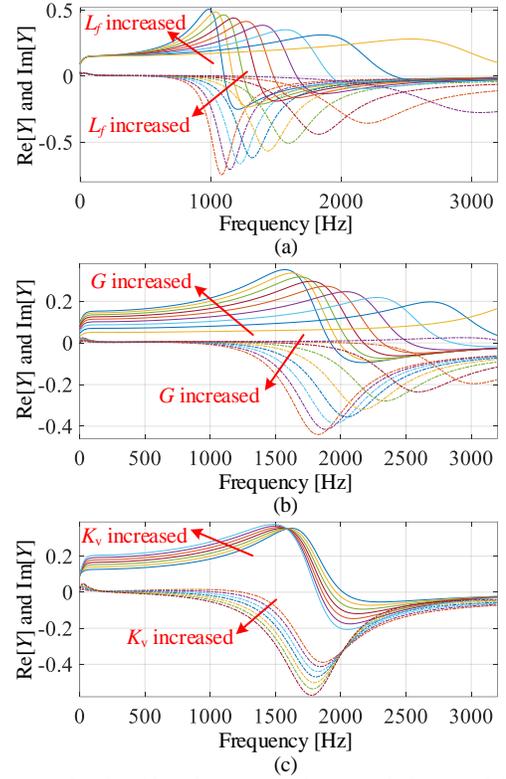

Fig. 7. Clusters of real and imaginary parts of output admittance of the proposed AD with parameters varying: (a) $L_f$, (b) $G$, and (c) $K_v$.

IV. CASE STUDY

A. System Description

A testing system with three inverters is used as case study to validate the proposed method of damping calculation and stability improvement. As shown in Fig. 8, three inverters that delivering the power from wind/solar are connected to the network, where the line impedances (i.e., $Z_{lin1}$ and $Z_{lin2}$) are considered. All these inverters are integrated to the grid through busbar transformer and cable.

Equivalent circuits of lines, cable, and transformer are given in Fig. 8, and inverter output admittance can be obtained by either mathematical modeling [7] or measurement technique [12]. Then the impedance-based equivalent circuit of the testing system can be derived. Parameters of inverters, lines, cable, and transformer can be found in tables II-III in Appendix.

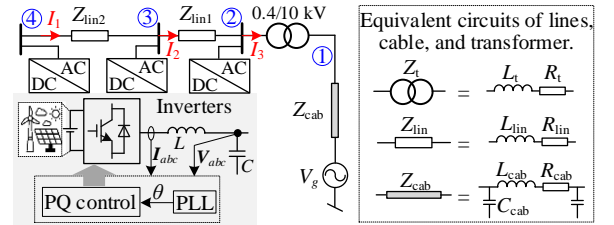

Fig. 8. Single-line diagram of testing case.

## B. Critical Eigenvalues and Stability Improvement by AD

According to parameters in tables II-III, the eigenvalues of $Y_{nod}$ are demonstrates in Fig. 9 (solid lines refer to real parts of eigenvalues, and dash lines refer to imaginary parts). It shows that three critical eigenvalues can be found, i.e., $\lambda_8$, $\lambda_5$, and $\lambda_6$, where three critical points, namely A ($f_{cr1}$=203Hz, Re[$\lambda_8$]=-0.0236), B ($f_{cr2}$=1821Hz, Re[$\lambda_5$]=-0.0049), and C ($f_{cr3}$=1957Hz, Re[$\lambda_6$]=-0.0074), should be concerned.

According to (7), each critical eigenvalue's compensation coefficient $K_C$ at different nodes can be obtained, as listed in table IV. It can be found that $K_C$ of $\lambda_8$ at Node 4 has the maximum value, while $K_C$ of $\lambda_5$ or $\lambda_6$ at Node 3 has the maximum value. It means that installing the AD at Node 4 is the most beneficial for $\lambda_8$, whereas at Node 3 is the most beneficial for $\lambda_5$ and $\lambda_6$. However, considering that the required damping compensation for $\lambda_8$ is larger than those for $\lambda_5$ and $\lambda_6$, it is highly recommended to install the proposed AD at Node 4 for efficient stability improvement.

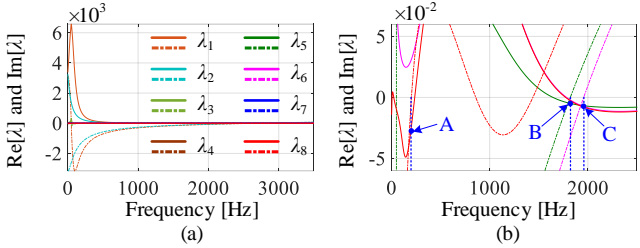

Fig. 9. Eigenvalues of $Y_{nod}$: (a) full view, and (b) zoom view of critical ones.

TABLE IV
COMPENSATION COEFFICIENT AT DIFFERENT NODES

| Nodes | $K_C$ of $\lambda_8$ at $f_{cr1}$ | $K_C$ of $\lambda_5$ at $f_{cr2}$ | $K_C$ of $\lambda_6$ at $f_{cr3}$ |
|---|---|---|---|
| Node 1 | 0.0000-0.0000$j$ | 0.0000-0.0000$j$ | 0.0000-0.0000$j$ |
| Node 2 | 0.0006-0.0000$j$ | 0.0023-0.0000$j$ | 0.0023-0.0000$j$ |
| Node 3 | 0.2322+0.0004$j$ | **0.7648-0.0000$j$** | **0.7648-0.0000$j$** |
| Node 4 | **0.7672-0.0003$j$** | 0.2329-0.0000$j$ | 0.2329-0.0000$j$ |

The required damping compensation $Y_{ad}$ for each critical eigenvalue can be calculated according to Algorithm 1. To provide sufficient stability margin for the case system (set $\varepsilon$=0.005 in this paper), Re[$Y_{ad}$] is supposed to be greater than 0.05 in the frequency range from 100 Hz to 2000 Hz when AD is installed at Node 4 (AD's parameters are listed in table I).

Fig. 10 shows critical eigenvalues with AD installed at Node 4. It can be seen that the real parts of critical eigenvalues $\lambda_8$, $\lambda_5$, and $\lambda_6$ can be improved by AD at their crossover frequencies.

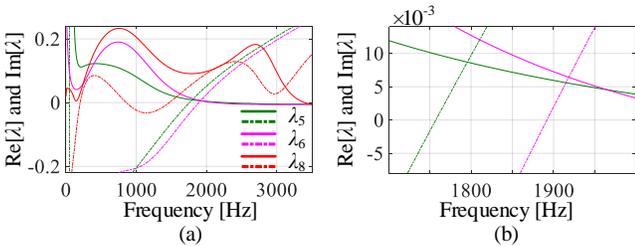

Fig. 10. Critical eigenvalues with the proposed AD installed at Node 4: (a) full view, and (b) zoom view of $\lambda_5$, and $\lambda_6$.

However, as demonstrated in Fig. 11, when AD installed at Node 3, the real parts of critical eigenvalues $\lambda_5$, and $\lambda_6$ can be improved, but the real part of critical eigenvalue $\lambda_8$ is still negative at crossover frequency, and perhaps larger damping compensation $Y_{ad}$ is required. This indicates that installation location of AD could be optimized according to $K_C$ for global stability improvement.

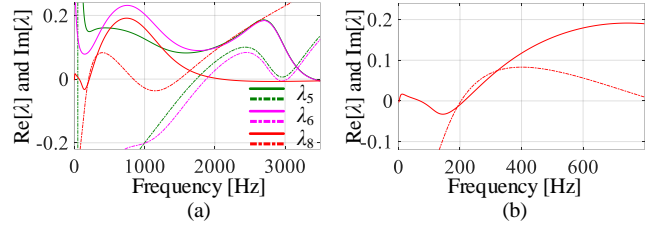

Fig. 11. Critical eigenvalues with the proposed AD installed at Node 3: (a) full view, and (b) zoom view of $\lambda_8$.

Fig. 12 illustrates the real and imaginary parts of $Y_{ad}$ of the traditional AD. It can be found that, without the proposed feedforward control strategy, i.e., $H_i$=0, the real part of $Y_{ad}$ of the traditional AD declines sharply as frequency increases. It means that the traditional AD can provide smaller Re[$Y_{ad}$] for damping compensation in high frequency range. Besides, it shows relatively large negative Im[$Y_{ad}$] in wide frequency range.

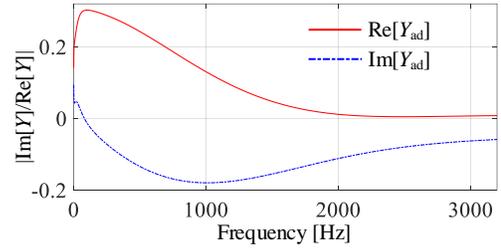

Fig. 12. Real and imaginary parts of $Y_{ad}$ of the traditional AD ($H_i$=0).

Fig. 13 illustrates critical eigenvalues with the traditional AD installed at Node 4. It can be seen that although the critical eigenvalue $\lambda_8$ is fully compensated in low frequency range, the critical eigenvalues $\lambda_5$ and $\lambda_6$ are still of inadequate compensation at their crossover frequencies. The reason is due to that the relatively large negative Im[$Y_{ad}$] makes the crossover frequencies of $\lambda_5$ and $\lambda_6$ have significant changes and shift to higher frequency range where no sufficient Re[$Y_{ad}$] for damping compensation. This implies that the proposed quasi-pure resistive AD presents better performance for damping compensation and global stability improvement.

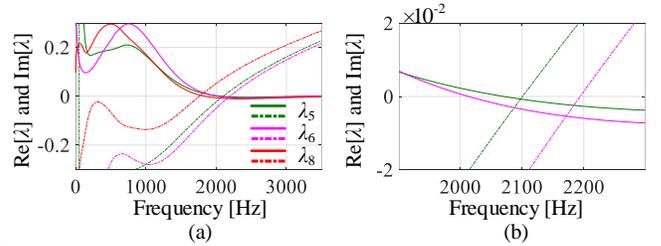

Fig. 13. Critical eigenvalues with the traditional AD installed at Node 4: (a) full view, and (b) zoom view of $\lambda_5$, and $\lambda_6$.

## C. Simulation Verification

To further verify the effectiveness of the proposed damping compensation method for stability improvement, simulation model of the case system is constructed in PSACD/EMTDC. Figs. 14-15 show simulation results of current waveforms with the proposed AD installed at Node 4 and 3, respectively ($I_1$, $I_2$, and $I_3$ can be found in Fig. 8, and $I_{ad}$ refers to output current of AD).

From Fig. 14, it can be seen that the system can be stable with AD installed at Node 4, but when removing AD at 0.8 s, the system becomes unstable and current waveforms appear severely distorted. Fig. 15 shows that the system cannot be stable with the proposed AD installed at Node 3, and when removing AD at 0.8 s, the system is still unstable.

Fig. 16 show simulation results of current waveforms with the traditional AD installed at Node 4. It can be found that the system cannot be stable with the traditional AD installed at Node 4. The simulation results are consistent with the theoretical analysis.

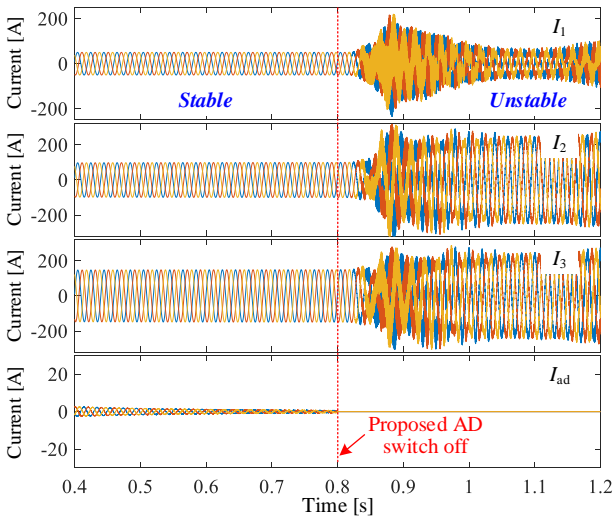

Fig. 14. Simulation results of current waveforms with the proposed AD installed at Node 4.

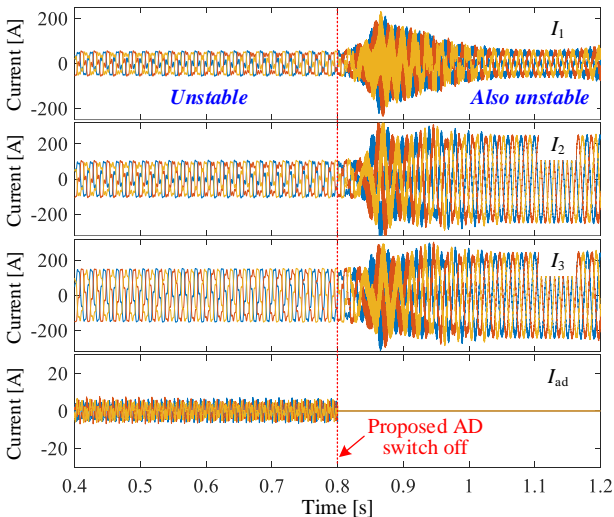

Fig. 15. Simulation results of current waveforms with the proposed AD installed at Node 3.

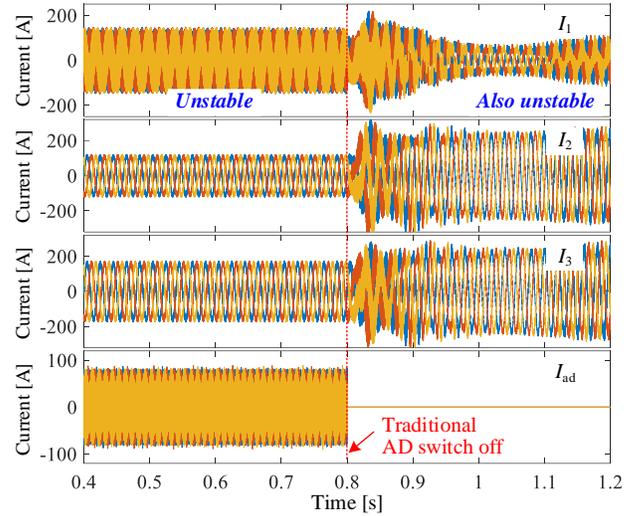

Fig. 16. Simulation results of current waveforms with the traditional AD installed at Node 4.

## V. EXPERIMENTAL VERIFICATION

To further verify the proposed method for global stability improvement of inverter-based systems, experiments of the testing system are carried out on the platform of the control hardware in the loop. Parameters in tables I-III are used in the experiments. Figs. 17-19 show experimental results.

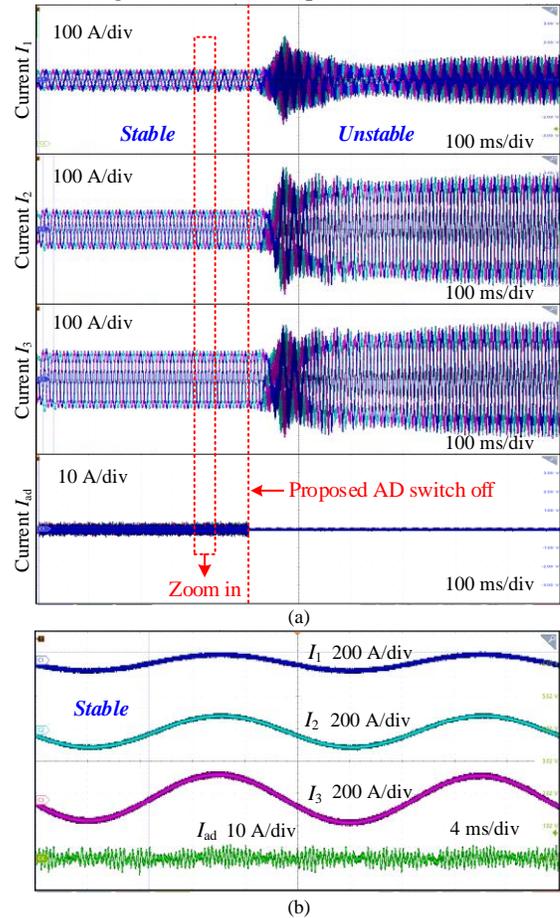

Fig. 17. Experimental results of current waveforms with the proposed AD installed at Node 4: (a) full view, and (b) zoom in view.

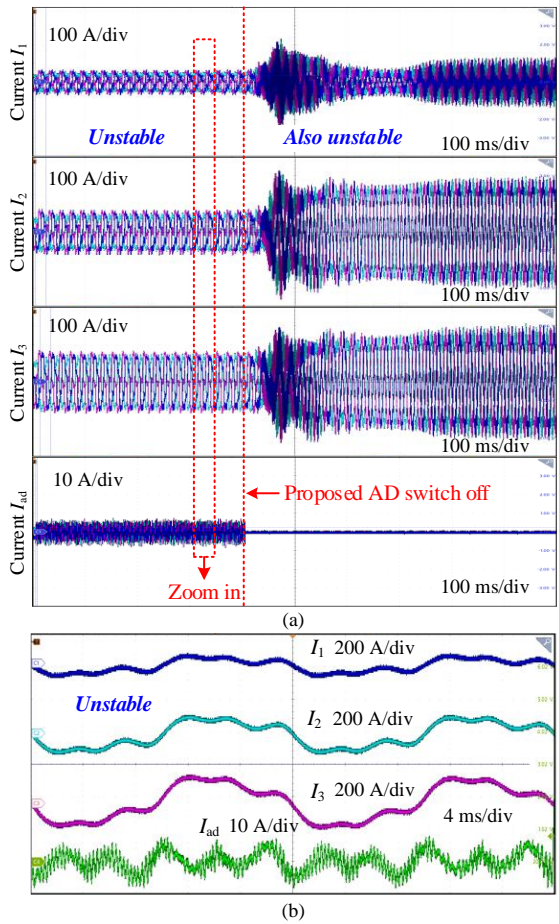

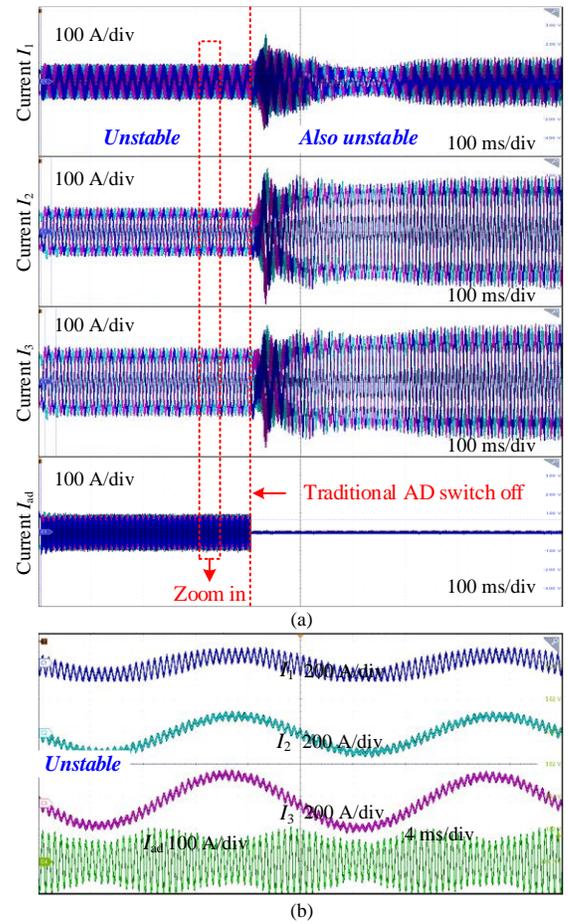

Fig. 18. Experimental results of current waveforms with the proposed AD installed at Node 3: (a) full view, and (b) zoom in view.

Fig. 19. Experimental results of current waveforms with the traditional AD installed at Node 4: (a) full view, and (b) zoom in view.

Figs. 17 and 18 depict experimental results of current waveforms with the proposed AD installed at Node 4 and at Node 3, respectively. It can be seen that the system can be stable with the proposed AD installed at Node 4, but becomes unstable when removing AD. By contrast, the system cannot be stable with the proposed AD installed at Node 3, let alone AD removed from Node 3. It indicates that the system global stability can be ensured when the AD is installed at Node 4, not Node 3, which is consistent with the theoretical analysis results shown in Figs. 10-11.

Fig. 19 depicts experimental results of current waveforms with the traditional AD installed at Node 4. The results show that that the system cannot maintain stable operation regardless of whether the traditional AD is connected or not, indicating that the damping compensation effect of the traditional AD for global stability improvement is inadequate.

## VI. CONCLUSION

To effectively address instability-induced oscillation issues and enhance global stability of inverter-based systems, firstly, this paper presents a damping calculation algorithm to quantitatively calculate the required damping compensation for global stability improvement, based on which the compensation coefficient of AD can be obtained and optimal location can be identified. Secondly, to adequately compensate negative damping of system in a wide frequency range, we propose a specific AD that has the feature of broadband quasi-pure resistor by using the output current feedforward control strategy. Compared with the traditional AD, the proposed quasi-pure resistive AD can effectively achieve broadband damping compensation in wide frequency range. And finally, a testing system with three inverters is used as case study, showing that the proposed method provides a promising solution to efficiently enhance global stability improvement of inverter-based systems.

APPENDIX

See tables I-III.

TABLE I
PARAMETERS OF AD

| Parameter | Value |
|---|---|
| DC-link voltage $V_{dc}$ (V) | 750 |
| Inductance of filter inductor $L$ (mH) | 0.8 |
| $D$ channel current $I_d$ (A) | 0 |
| $Q$ channel current $I_q$ (A) | 0 |
| Proportional gain of current controller $k_{pi}$ | 5 |
| Integral gain of current controller $k_{ii}$ (s$^{-1}$) | 100 |
| Damping ratio of notch filter $\xi$ | 0.707 |
| Time constant of lag filter $\tau$ | 0.0014 |
| Ratio of zero locations to pole locations of lag filter $\beta$ | 2 |
| Cut-off frequency of low pass filter $\omega_{low}$ (rad/s) | 12566.36 |
| Cut-off frequency of the intended low pass filter $\omega_c$ (rad/s) | 21991.13 |
| Gain of the intended low pass filter $G$ | 0.06 |
| Sampling frequency $f_s$ (kHz) | 40 |

TABLE II
PARAMETERS OF INVERTERS

| Parameter | Value |
|---|---|
| DC-link voltage $V_{dc}$ (V) | 750 |
| Inductance of filter inductor $L$ (mH) | 2.5 |
| Capacitance of filter capacitor $C$ (μF) | 15 |
| $D$ channel current $I_d$ (A) | 50 |
| $Q$ channel current $I_q$ (A) | 0 |
| Proportional gain of current controller $k_{pi}$ | 10 |
| Integral gain of current controller $k_{ii}$ (s$^{-1}$) | 300 |
| Proportional gain of PLL $k_{pPLL}$ | 6 |
| Integral gain of PLL $k_{iPLL}$ (s$^{-1}$) | 100 |
| Sampling frequency $f_s$ (kHz) | 10 |

TABLE III
PARAMETERS OF LINES, CABLE, AND TRANSFORMER

| Parameter | Value | Parameter | Value | Parameter | Value |
|---|---|---|---|---|---|
| $L_{lin1}$ (mH) | 1.5 | $R_{lin2}$ (Ω) | 0.06 | $L_{cab}$ (mH) | 0.3 |
| $R_{lin1}$ (Ω) | 0.04 | †$L_t$ (mH) | 0.0764 | $R_{cab}$ (Ω) | 0.2 |
| $L_{lin2}$ (mH) | 2.0 | †$R_t$ (Ω) | 0.0032 | $C_{cab}$ (μF) | 12 |

†Transformer parameters referred to its low-voltage side.